\documentclass[aps,prd,preprint,floats,superscriptaddress,nofootinbib]{revtex4}
\usepackage[utf8]{inputenc}
\usepackage{amsmath,amssymb}
\usepackage{graphicx}
\usepackage{color,slashed}
\pdfoutput=1

\begin{document}
\title{\bf Ultra-high-energy neutrino scattering in an anomalous $U(1)$ effective field theory}

\author{Chuan-Hung Chen}
\email[E-mail: ]{physchen@mail.ncku.edu.tw}
\affiliation{Department of Physics, National Cheng-Kung University, Tainan 70101, Taiwan}
\affiliation{Physics Division, National Center for Theoretical Sciences, Taipei 10617, Taiwan}

\author{Cheng-Wei Chiang}
\email[E-mail: ]{chengwei@phys.ntu.edu.tw}
\affiliation{Department of Physics and Center for Theoretical Physics, National Taiwan University, Taipei 10617, Taiwan}
\affiliation{Physics Division, National Center for Theoretical Sciences, Taipei 10617, Taiwan}

\author{Chun-Wei Su}
\email[E-mail: ]{r10222026@ntu.edu.tw}
\affiliation{Department of Physics and Center for Theoretical Physics, National Taiwan University, Taipei 10617, Taiwan}

\date{\today}

\begin{abstract}
A unique characteristic of exponentially growing scattering amplitude arises in an anomalous Abelian effective field theory when an extremely light Dirac neutrino mass is introduced to break the symmetry. We show that the low energy effective Lagrangian can be made explicitly gauge invariant with the help of a nonlinear representation of the Goldstone or Stueckelberg field.  We study the peculiar feature of exponential growth in the ultra-high-energy neutrino-nucleon inelastic scattering.  It is found that the inelastic scattering cross section is highly sensitive to the ratio of gauge coupling to the gauge boson mass, $g_X/m_X$.  When the IceCube measurement of ultra-high-energy neutrinos, which is consistent with the standard model prediction up to $E_\nu \sim 6$ PeV, is taken into account, the inferred constraint on $g_X/m_X$ is more severe than that obtained from the events of mono-lepton$+$missing transverse energy at the LHC.  A muon collider with a collision energy of $10$ TeV can be a good environment other than hadron colliders to probe the novel effect. 

\end{abstract}
\maketitle

\section{Introduction}

A light $U(1)$ gauge boson is an interesting subject that has been proposed to resolve some anomalies indicated by experimental observations~\cite{Gninenko:2001hx,Pospelov:2008zw,Feng:2016ysn}.  It was found that such a light gauge boson could arise from an anomalous $U(1)$ effective field theory (EFT) at low energies, which could not only be derived from an ultraviolet (UV) complete theory, but also be consistently quantized in perturbation theory~\cite{Krasnikov:1985bn,Faddeev:1986pc,Preskill:1990fr}.  Intriguingly, several characteristic scales between low energies and UV completion may appear, depending on what roles the heavy degrees of freedom play at high energies. 

It has been shown that when gauge anomaly cancellation is achieved by introducing new heavy chiral fermions, whose chiral components generally carry different $U(1)$ charges, gauge invariance will be broken by the heavy fermion mass terms.  However, the gauge invariance can be restored if a scalar field $\theta$ in a nonlinear representation of the $U(1)$ symmetry is introduced to remove the gauge phase factor from the heavy fermion~\cite{Craig:2019zkf}, where the scalar field $\theta$ is a Goldstone boson acting as the longitudinal mode of the massive $U(1)$ gauge boson.  As a consequence, the interaction between $\theta$ and the fermions is nonrenormalizable.  Such a gauge-invariant, anomaly-free EFT is expected to be valid up to a scale given by the ratio between the $U(1)$ gauge boson mass and the charge difference between the left- and right-chiral fermions.  Interestingly, the nonrenormalizable interaction would lead to a novel multi-longitudinal gauge boson coupling to the heavy fermions, and the associated scattering amplitude features an exponential growth with energy~\cite{Craig:2019zkf}. 

Inspired by the above-mentioned peculiar property, the authors in Ref.~\cite{Ekhterachian:2021rkx} applied the anomalous EFT and nonrenormalizable interaction to the active light neutrino system.  In addition to the severe bound on the cutoff scale from the unitarity condition, it was found that the constraint on the ratio of the gauge coupling to gauge boson mass, $g_X/m_X$, was stronger when the involved energy of the system got higher.  For instance, the bound from $pp\to W^* \to \ell \bar\nu$ at $2$~TeV is severer than that from the $W$ decay at the $m_W$ scale.  By the same token, we anticipate that the cosmic ultra high-energy (UHE) neutrino scattering off matters can be a good place to explore the effects of the anomalous $U(1)$ EFT. 

It has been known for a long time that the Glashow resonance~\cite{Glashow:1960zz} can be achieved if the incident antielectron neutrino energy reaches $E_\nu = m^2_W/(2 m_e) \approx 6.3$~PeV.  Evidence of the Glashow resonance is now reported by the IceCube neutrino observatory~\cite{IceCube:2021rpz}.  To match the constraint from the process $\bar \nu e \to \bar\nu\, \theta^n e$ to that from $pp\to W^* \to \ell \bar\nu$ at $2$~TeV, the energy of the incident UHE antielectron neutrino has to satisfy $E_\nu \gtrsim 4~ \text{TeV}^2/(2 m_e) \approx 3.9$~EeV, which is far above the current energy that IceCube can probe.  However, if we change the target from an electron to a nucleon with the mass of $\sim 1$~GeV,  the center-of-mass energy of the $\nu$-$N$ system is $\sqrt{s} \gtrsim 4.47$~TeV when the incident neutrino energy $E_\nu \geq 10$~PeV.  Based on such an estimate, we expect that the nonrenormalizable interaction can be better probed or constrained by the UHE neutrino-nucleon inelastic scattering.  In this study, we therefore focus the analysis on the $\nu N \to \nu \, \theta^n X$ process.  

We find that the UHE neutrino-nucleon inelastic scattering has the following features: (i) due to the exponential growth feature of the cross section, the $\nu N\to \nu \, \theta^n X$ process is insensitive to how to formulate the hadronic effects; (ii) the dominant regions of the Bjorken $x$ and $y$ variables are at $x\sim O(1)$ and $y\ll 1$, which are different from the $\nu N \to \nu X$ process dominated in the small $x$ region; (iii) the cross section is highly sensitive to $g_X/m_X$; and (iv) the constraints from $\sigma(\nu N\to \nu \,  X)$ with $E_\nu \geq 10$~PeV are severer than those from the direct measurements at the LHC.  

In the rest of the paper, we first review in Section~\ref{sec:Anomalous U(1) effective theory} the gauge-invariant, anomaly-free $U(1)$ model that leads to a desired anomalous $U(1)$ EFT at low energies, with the feature of having nonrenormalizable interactions between neutrinos and multiple Goldstone bosons.  Dirac-type neutrino mass is generated by the nonrenormalizable effects, which can originate from a UV complete theory, with the Abelian Higgs model as the simplest implementation.  Based on the anomalous $U(1)$ EFT, we derive in Section~\ref{sec:UHE neutrino-nucleon deep inelastic scattering} the scattering cross section for the process $\nu N \to \nu \, \theta^n X$ and make a detailed analysis.  As a comparison, we also show the standard model predictions for the $\nu N \to \nu X$ inelastic scattering.  Finally, we summarize what we learn from the study in Section~\ref{sec:Summary}.  

\section{Anomalous $U(1)$ effective theory \label{sec:Anomalous U(1) effective theory}}

To investigate the peculiar feature of neutrino coupling to a light massive gauge boson in an anomalous $U(1)$ gauge theory, we start from a specific model and establish a partially UV-completed Abelian EFT, where the EFT is $U(1)$ anomaly-free and the neutrino mass arises from higher dimensional operators, controlled by a cutoff scale $\Lambda'$.  Since the primary purpose of this study is to examine the implications of an anomalous $U(1)$ EFT at low energies, we refer the readers to Ref.~\cite{Craig:2019zkf} for a fully renormalizable theory above the $\Lambda'$ scale. 


In the following, we start by constructing a model that is $U(1)$ gauge anomaly-free at the $\Lambda'$ scale.  We assume that the particles in the standard model (SM) do not carry the new $U(1)$ charge.  For the purpose of simplicity, we focus on Dirac-type neutrinos and, thus, a light right-handed neutrino carrying the $U(1)$ charge is introduced.  Since the main effect is insensitive to the number of neutrino species, we just concentrate on one neutrino flavor in the analysis, and it should be straightforward to extend the analysis to the case of three flavors~\cite{Ekhterachian:2021rkx}.  We note that in an alternative scheme, the $U(1)$ gauge boson can be assumed to only couple to the left-handed neutrino in a way that the charged lepton interacting with the $U(1)$ gauge boson is a vector-like coupling~\cite{Craig:2019zkf,Ekhterachian:2021rkx}.  In this case, the gauge boson can contribute to lepton-dependent processes, such as lepton $g-2$.  Since our focus is on showing the peculiar effect of an $U(1)$ anomalous EFT, we take the simplest extension of the SM by assuming that the new $U(1)$ gauge boson only couples to new particles beyond the SM.
 
Although there is a $U(1)$ gauge anomaly in the model at low energies, it can be cancelled if heavy SM singlet chiral fermions with properly chosen $U(1)$ charges appear at high energies.  In addition, we add a complex scalar $\Phi$ to the model, so that the mass of heavy singlet fermions can be generated via spontaneous symmetry breaking and the $U(1)$ gauge invariance can be preserved, though it is broken by anomaly at the scale below the heavy fermion mass.  The $U(1)$ gauge-invariant and gauge anomaly-free Lagrangian is given by:
\begin{align}
  \label{eq:L0}
  {\cal L} =& 
  - \frac{1}{4} F_{\mu\nu} F^{\mu\nu} + (D_\mu \Phi)^\dagger (D^\mu \Phi) 
  - \frac{\lambda}{2} \left(|\Phi|^2 - \frac{f^2}{2} \right)^2 
  + \bar N i \slashed{D} N + \bar\nu_R i \slashed{D} \nu_R 
  \nonumber \\
  & - \left(y_N \Phi^* \bar{N}_L N_R + y_\nu \left( \frac{\Phi}{\Lambda'}\right)^q \bar L \tilde H \nu_R + \mbox{H.c.} \right)\,,
\end{align}
where $F_{\mu\nu}$ is the gauge field strength tensor associated with the new $U(1)$ gauge field $X^\mu$, $\tilde H = i\tau_2 H^*$ with $H$ being the SM Higgs doublet, $L$ is the SM lepton doublet, and $f$ is the vacuum expectation value (VEV) of $\Phi$.  The $U(1)$ charges of $N_{R,L}$, $\Phi$, and $\nu_R$ are respectively assigned to be $Q^N_{R,L}$, $Q_\Phi=Q^N_R-Q^N_L$, and $Q^\nu_R= - q Q_\Phi$, 
with the assumption that $Q^N_{L} \neq Q^N_{R}$.  We note that since our purpose is to illustrate the possible anomaly free gauge model at the $\Lambda$ scale, the SM Higgs-related terms, responsible for the electroweak symmetry breaking as in the SM, are skipped.  When the values of $Q^N_{R,L}$ are fixed, $q$ can be determined by the $U(1)^3_X$ anomaly cancellation, which is formulated by $-q^3 Q^3_\Phi + (Q^N_R)^3- (Q^N_L)^3=0$~\cite{Craig:2019zkf}.  With $Q^N_R > Q^N_L$, the relation of $q$ and $Q^N_{R,L}$ is obtained as:
 \begin{equation}
 q= \frac{\left[(Q^N_R)^3 - (Q^N_L)^3 \right]^{1/3}}{Q^N_R - Q^N_L}\,.
 \end{equation}
The covariant derivative is defined as:
 \begin{equation}
 %
 %
 D^\mu F  = (\partial^\mu - i g_0 Q_F X^\mu ) F
 ~,
 \end{equation}
where $g_0$ is the $U(1)$ gauge coupling and $Q_F$ is the $U(1)$ charge of the corresponding particle $F$.  The nonrenormalizable $y_\nu$ term is responsible for the generation of neutrino mass, suppressed by powers of $f/\Lambda'$.

In order to show the gauge invariance after integrating out the radial mode of the scalar field $\Phi$, we parametrize it as:
   \begin{equation}
   \Phi= \frac{f + \rho}{\sqrt{2}} e^{i \frac{\theta}{f}}\,,
   \end{equation}
where $\rho$ is the real radial mode and $\theta$ is the Goldstone field.  As a result, the anomalous effective Lagrangian with massive gauge boson can be obtained as:
 \begin{align}
 \label{eq:Leff}
 {\cal L}_{\rm eff}  & = - \frac{1}{4} F_{\mu \nu} F^{\mu\nu} + \frac{1}{2}  \left( m_X  X^\mu -  \partial^\mu \theta \right)^2  + \bar\nu_R i \slashed{D} \nu_R  -\left(    e^{i q \frac{ \theta}{f}} \bar\nu_L m_\nu  \nu_R + \mbox{H.c.} \right)\,,
 \end{align} 
where $m_X = g_0 Q_\Phi f$,  $m_\nu=(f/\sqrt{2} \Lambda')^q  y_\nu v_H/\sqrt{2}$, and $v_H$ is the VEV of the Higgs doublet field $H$.  It is then easy to verify that the Lagrangian in Eq.~(\ref{eq:Leff}) is invariant under the gauge transformations:
  \begin{align}
   X^\mu & \longrightarrow X^\mu + \partial^\mu \alpha\,, \nonumber \\
   \theta & \longrightarrow \theta + m_X \alpha\,, \nonumber \\
  %
  %
   \nu_R & \longrightarrow e^{ -i g_X \alpha} \nu_R\,,
  \end{align}
with $g_X \equiv - g_0 Q^\nu_R= q m_X/f $ and $\alpha (x)$ being the gauge function.  In this work, we focus on the scenario of having a gauge boson whose mass is parametrically small ($g_0 Q_\Phi \ll 1$) in comparison with the characteristic energy scales of the physical processes in consideration~\cite{Craig:2019zkf}. Since $m_\nu$, $m_X$, and $q$ are strongly correlated to the values of $y_\nu$, $g_0$, and $Q^N_{R,L}$ in the model, for the purpose of illustration, we take some benchmark values of these parameters to show the numerical results in Table~\ref{tab:para_values}, where  $f=\Lambda'=1$~TeV is applied. From the results in the table, it can be seen that $q$ strongly depends on $Q^N_R-Q^N_L$, and the value of $y_\nu$ leading to $m_\nu\sim O(10^{-2})$ eV is highly sensitive to $q$. In order to get small neutrino mass,  in addition to $g_0$, $m_X$ depends on not only $Q^N_R-Q^N_L$ but also the individual values of $Q^N_{R,L}$. 
\begin{table}[htp]
\caption{Numerical values of $m_\nu$, $m_X$, and $g_X/m_X$ based on some benchmarks of $g_0$, $Q^N_{R,N}$, and $y_\nu$ with $f=\Lambda'=1$ TeV. }
\begin{center}
\begin{tabular}{cccccccc} \\ \hline \hline
~~$g_0$~~ & ~~$Q^N_R$~~ & ~~$Q^N_L$~~ & ~~~ $y_{\nu}$~~~& ~~~ $q$~~~ &  ~~$m_\nu/{\rm eV}$~~  & ~~$m_X/{\rm GeV}$~~ & ~~$g_X/m_X$ \\ \hline 
$10^{-2}$ &  $1$ & $0.998$  & $\sqrt{4\pi}$&  $90.796$ & $0.0133$ & $0.02$   & $0.091$ \\ \hline 
$10^{-2}$ & $5$ &  $4.990$ & $\sqrt{4\pi}$ & $90.796$ & $0.0133$ & $0.10$ & $0.091$ \\ \hline
$10^{-3}$ & $1$ & $0.995$ & $10^{-5}$  & $49.242$ &  $0.067$ & $0.005$ & $0.049$ \\ \hline
$10^{-3}$ & $5$ & $4.975$ & $10^{-5}$  & $49.242$  & $0.067$ & $0.025$ & $0.049$ \\ \hline \hline

\end{tabular}
\end{center}
\label{tab:para_values}
\end{table}%

Since $\theta$ is the Goldstone boson and represents the longitudinal component of $X^\mu$, the last term in Eq.~(\ref{eq:Leff}) gives rise to the interaction between neutrino and the longitudinal component of $X^\mu$ at high energies using the equivalence theorem.  With the expansion
  \begin{equation}
  \label{eq:Int_nu_nphi}
  e^{i q \frac{ \theta}{f}} \bar\nu_L  m_\nu \nu_R 
  = \sum_{n} \bar \nu_L  \frac{m_\nu}{n!}\left( \frac{g_X \theta }{m_X }\right)^n \nu_R\,,
  \end{equation}
it is seen that the neutrinos can now have interactions involving the emission/absorption of multiple longitudinal gauge bosons.  A similar expansion can also be applied to the heavy singlet fermion.  It is found that the neutrino scattering amplitude involving multiple longitudinal gauge bosons has the novel feature of exponential growth in energy~\cite{Craig:2019zkf,Ekhterachian:2021rkx}.  In this paper, we study its effects on the UHE neutrino-nucleon inelastic scattering. 

Although the effective Lagrangian in Eq.~(\ref{eq:Leff}) originates from a specific gauge anomaly-free theory shown in Eq.~(\ref{eq:L0}), it is in principle not crucial to specify the anomaly-free theory when one studies the low-energy phenomena.  Basically, one can write down the effective Lagrangian based on gauge invariance and take the Lagrangian as an anomalous EFT at low energies.  Thus, $m_X$, $g_X$, and $m_\nu$ can be taken as free parameters and can be constrained or determined by experimental data.  Moreover, the cutoff scale of the anomalous EFT, which signals the breakdown of the perturbation theory and indicates the emergence of new degrees of freedom, can be determined through the unitarity requirement~\cite{Preskill:1990fr,Craig:2019zkf,Ekhterachian:2021rkx}. 

A low-energy EFT with a massive gauge boson and $U(1)$ gauge invariance can be achieved using the Stueckelberg mechanism, where a Stueckelberg scalar field is introduced to retain the gauge invariance~\cite{Stueckelberg:1938hvi,Ruegg:2003ps}. The massive gauge field theory with the Stueckelberg mechanism is renormalizable and unitary~\cite{Lowenstein:1972pr,Ruegg:2003ps}.  When a Dirac neutrino mass term $\bar\nu_L \nu_R$ is included in the model to break the $U(1)$ gauge symmetry, the Lagrangian still keeps a gauge-invariant form due to the Stueckelberg field.  Nevertheless, this is done at the cost of losing renormalizability.  Under the gauge transformations:
  \begin{align}
  X^\mu \rightarrow& X^\mu +  \partial^\mu \alpha \,, ~  B  \rightarrow B + m_X \alpha \,, ~
  \nu_R  \rightarrow  e^{-i g_X \alpha} \nu_R \,,
  \end{align}
the associated Stueckelberg Lagrangian can be written as:
\begin{align}
 {\cal L}_{\rm St} =& 
 -\frac{1}{4} F_{\mu \nu} F^{\mu \nu}  + \frac{m^2_X}{2} \left( X^\mu - \frac{1}{m_X}\ \partial^\mu B \right)^2 
 - \frac{1}{2 \beta } \left( \partial_\mu X^\mu + m_X \beta B \right)^2
 \nonumber \\
 & + \bar\nu_R i \slashed {D} \nu_R + m_\nu \left( e^{i \frac{g_X}{m_X} B} \bar \nu_L \nu_R + \mbox{H.c.} \right)\,,
 \label{eq:St}
\end{align}
where $B(x)$ is the Stueckelberg field to become the longitudinal component of $X^\mu$, and the term involving $\beta$ is the Lagrangian multiplier for gauge fixing.  It can be seen that the Stueckelberg field plays the same role as the Goldstone boson in Eq.~(\ref{eq:Leff}).  Using a similar expansion shown in Eq.~(\ref{eq:Int_nu_nphi}), we also obtain the nonrenormalizable neutrino couplings to multiple longitudinal gauge bosons. 

\begin{figure}[phtb]
\begin{center}
\includegraphics[scale=0.55]{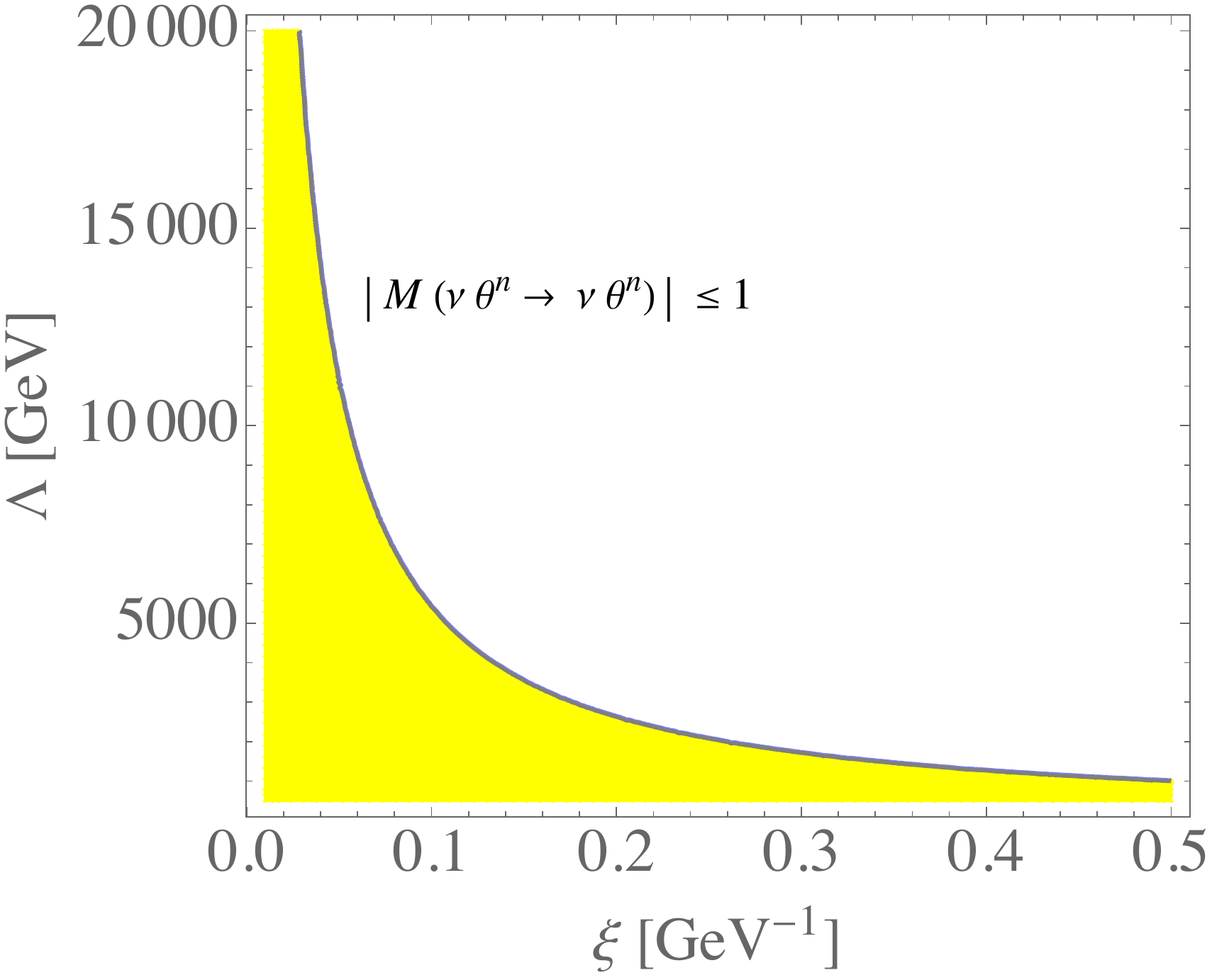}
\caption{ Unitarity bound on $\Lambda$ from the $\nu \, \theta^n \to \nu\, \theta^n$ scattering as a function of $\xi = g_X/m_X$. }
\label{fig:Uni_bound}
\end{center}
\end{figure}

The following analysis is based on the notation used in Eq.~(\ref{eq:Leff}).  The unitarity bound for the $\nu\, \theta^n \to \nu\, \theta^n$ scattering has been studied in Ref.~\cite{Ekhterachian:2021rkx}, and the scattering amplitude with $E \gg n m_X$ in the large $n$ limit is obtained as:
 \begin{equation}
 \label{eq:M_nu_nphi}
 M(\nu\, \theta^n \to \nu \, \theta^n) \approx \frac{m_\nu \xi}{2}  \frac{e^{3 (E \xi/4\pi)^{2/3}}}{2\sqrt{2} \pi^{3/2} (E \xi/4\pi)^2} \,,
 \end{equation}
where $\xi \equiv g_X/m_X$, and $E=\sqrt{P^2}$ is the center-of-mass (c.m.) energy of $\nu$-$\theta^n$ system.  It is clearly seen that the amplitude of the neutrino scattering with $n$ longitudinal modes of $X^\mu$ in the anomalous EFT increases exponentially in $g_X E/m_X$; that is, the bound is very sensitive to $g_X E/m_X$.  Taking $E$ as the cutoff scale $\Lambda$ of the EFT, we plot the contour for the unitarity condition of $|M(\nu\, \theta^n \to \nu \, \theta^n)| \leq 1$ in the plane of $\Lambda$ and $\xi$ in Fig.~\ref{fig:Uni_bound}, where $m_\nu=0.05$~eV is assumed.  It can be seen that when the cutoff scale $\Lambda$ increases, the upper bound on $\xi$ decreases.

\section{UHE neutrino-nucleon deep inelastic scattering \label{sec:UHE neutrino-nucleon deep inelastic scattering}}

The exponentially growing amplitude of $\nu\, \theta^n \to \nu\, \theta^n$, which arises from the nonrenormalizable interaction in Eq.~(\ref{eq:Int_nu_nphi}), indicates that the processes with  multi-longitudinal gauge-boson emissions in the final state can become significant when the c.m. energy in the system gets higher than the TeV scale, which is assumed to be orders of magnitude higher than the mass of the light $U(1)$ gauge boson.  For the purpose of illustration, it was shown in Ref.~\cite{Ekhterachian:2021rkx} that the constraint on the parameter $\xi$ from $pp\to W^* \to \ell \bar\nu$ at $2$~TeV scale is much severer than that from the decay width of $W$ boson at the $m_W$ scale.
\footnote{We note in passing that the bound on $\xi$ from the $Z$ boson decay is similar to that from the $W$ decay, reflecting the fact that the bound is less sensitive to a small change in the energy scale involved in the physical processes.}  In the following analysis, we investigate the implications of the nonrenormalizable effective interactions on the UHE neutrino-nucleon inelastic scattering. 

Since the UHE neutrino-nucleon inelastic scattering process involves the $Z$ mediation, here we list the $Z$-boson interactions with neutrinos and quarks in the SM as follows:
 \begin{equation}
 {\cal L}_{Z \bar f f}  = - \frac{g}{2 c_W} \bar f \gamma_\mu (g^f_R P_R + g^f_L P_L ) f Z^\mu\,,
 \end{equation}
where $g$ is the $SU(2)_L$ gauge coupling, $P_{L,R}$ are the chiral projection operators, $c_W~(s_W)=\cos\theta_W~ (\sin\theta_W)$, and the couplings to fermions are given by:
  \begin{align}
  \label{eq:Zint_SM}
  g^{u}_R & = -\frac{4}{3} s^2_W\,,~ g^{u}_L = 1- \frac{4}{3} s^2_W\,, \nonumber \\
   g^{d}_R & = \frac{2}{3} s^2_W\,,~ g^{d}_L = -1+ \frac{2}{3} s^2_W \,, 
  \end{align}
and $g^{\nu}_{L(R)}=1 (0)$.

\subsection{cross section for $\nu N\to \nu\, \theta^n X$}

\begin{figure}[phtb]
\begin{center}
\includegraphics[scale=0.5]{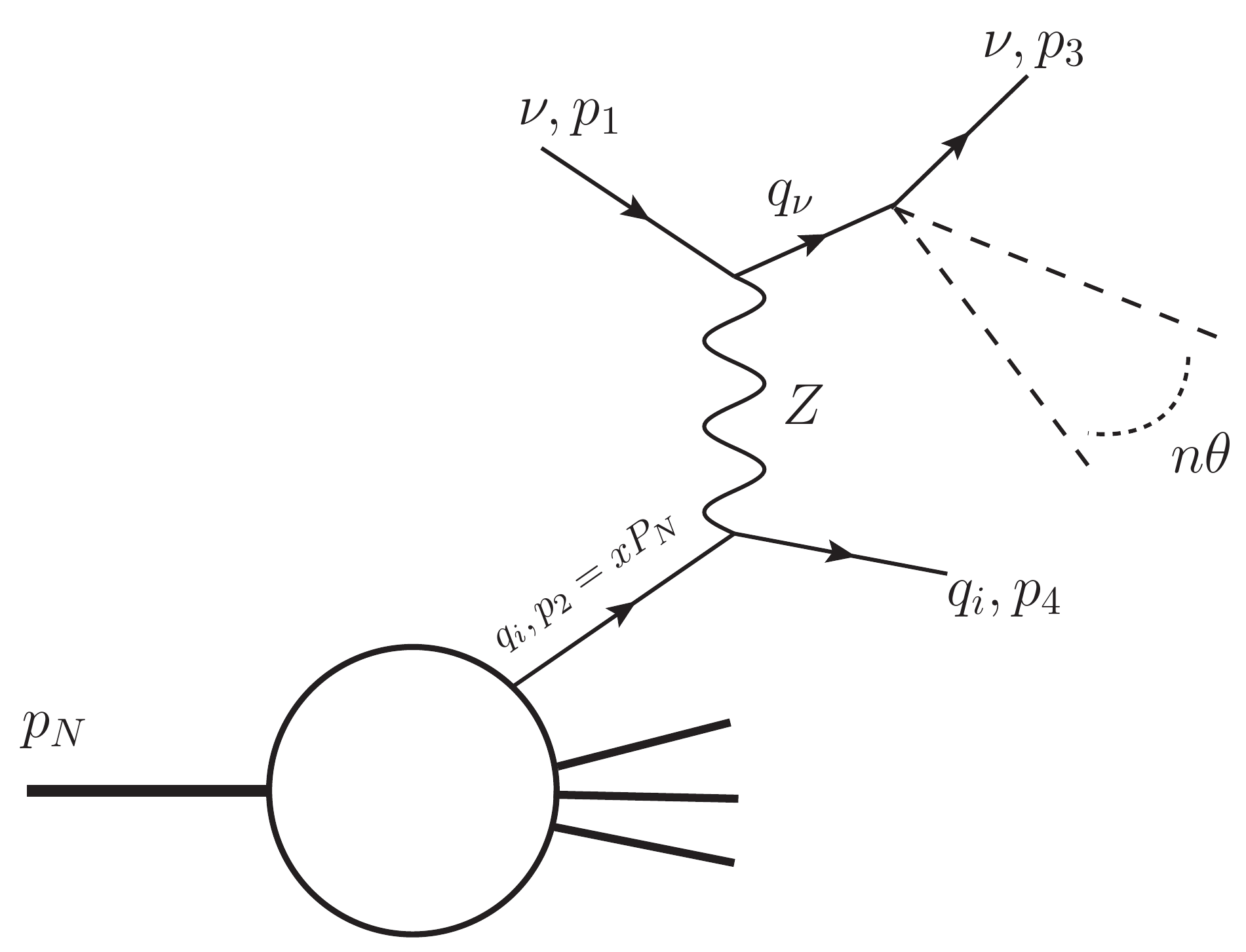}
\caption{ A sketch of $\nu N \to \nu \, \theta^n  X$ scattering.}
\label{fig:NuDis}
\end{center}
\end{figure}

The Feynman diagram for the $\nu N \to \nu\, \theta^n X$ inelastic scattering is shown in Fig.~\ref{fig:NuDis}.  In order to formulate the scattering cross section, as in the case of $\nu N\to \nu X$ inelastic scattering, we define the relevant kinematic variables:
 \begin{align}
 Q^2 = - (p_1 - q_\nu)^2\,,~
  x = \frac{Q^2}{ 2 p_N\cdot (p_1- q_\nu)}\,,~ 
   y = \frac{p_N \cdot(p_1-q_\nu)}{p_N\cdot p_1}\,,~ 
   z = \frac{q_\nu^2}{s} \,,
   \label{eq:variable definitions}
 \end{align}
where $p_i$ are the on-shell particle momenta, $q_\nu$ is the off-shell neutrino momentum, $Q^2$ is the momentum transfer squared, $x$ and $y$ are the Bjorken variables, and $s=2 m_N E_\nu$ is the c.m. energy of the $\nu$-$N$ system.  Using the neutrino couplings to $n\theta$ given in Eq.~(\ref{eq:Int_nu_nphi}) and the $Z$ boson couplings to quark and neutrino given in Eq.~(\ref{eq:Zint_SM}), the differential cross section for $\nu q \to \nu \,\theta^n q$ as a function of $y$ and $z$ is obtained as:
\begin{align}
\label{eq:xsq1}
\frac{d^2 \sigma_{\nu q \to \nu \,\theta^n q}}{dy dz} 
=& 
\frac{m^2_\nu m^4_Z G^2_F}{ 8 \pi  (n-1)! n! (n+1)!}  \left(  \frac{\xi \sqrt{s}}{4 \pi}\right)^{2n} \frac{z^{n-2}}{(Q^2 + m^2_Z)^2}   \nonumber \\
& \times \left[ |g^{q}_L|^2 ( x-z) + |g^q_R|^2 (1-y) \left(x(1-y) -z \right) \right] \,, 
\end{align}
with $Q^2=s x y$.  When deriving the above differential cross section, we have taken $\nu$ and $\theta$ to be massless due to $s \gg n m_X, m_\nu$, and the $(n+1)$-body phase space for $\nu+n\theta$ in the final state is given by~\cite{Abu-Ajamieh:2020yqi}:
 \begin{equation}
\int (2\pi)^4 \delta^4\left(P-\sum p_k \right)\frac{d^3 p_1}{(2\pi)^3 2 E_1} ...\frac{d^3p_{n+1}}{(2\pi)^3 2E_{n+1}}=\frac{1}{8\pi n! (n-1)!}  \left( \frac{\sqrt{P^2}}{4\pi}\right)^{2n-2}\,.
\end{equation}
Summing over the number of scalars for $n\geq 2$, we obtain:
\begin{align}
\label{eq:szLR}
 s_{zL} (x,y,z)&\equiv \sum_{n=2}  \frac{a^{2n} z^{n-2} (x-z)}{(n-1)! n! (n+1)!} =\frac{a^2  (x-z)}{2 z } \left[ _0F_2(;2,3;a^2 z)-1\right]\,, \nonumber \\
 s_{zR}(x,y,z)&\equiv \sum_{n=2} \frac{a^{2n} z^{n-2} (1-y) (x(1-y)-z)}{(n-1)! n! (n+1)!}  \nonumber \\
 &=\frac{a^2  (1-y) \left( x(1-y) - z \right)}{2 z } \left[ _0F_2(;2,3;a^2 z)-1\right] \,,
\end{align}
where we define $a \equiv \xi \sqrt{s}/(4\pi)$.

When we integrate the $z$ variable in the region of $z=[0,x (1-y)]$, the differential cross section in Eq.~(\ref{eq:xsq1}) becomes:
\begin{align}
\label{eq:xsq}
\frac{d \sigma_{\nu q \to \nu \, \theta^n \, q}}{dy} 
=& 
\frac{m^2_\nu m^4_Z G^2_F}{ 8 \pi (Q^2 + m^2_Z)^2}   \left[ |g^{q}_L|^2 s_L(x,y) + |g^q_R|^2 s_R(x,y) \right] \,,  \nonumber\\
s_L(x,y) 
=& \frac{ a^4 x^2 (1-y) }{24}  [ 2 y\, _2F_4(1,1;2,2,3,4; a^2 x(1-y)) \nonumber \\
& + (1-y)\, _2F_4(1,1;2,3,3,4; a^2 x(1-y)) ]\,, \nonumber \\
s_R (x,y)
=& 
\frac{ a^4 x^2 (1-y)^3 }{24}\, _2F_4(1,1;2,3,3,4; a^2 x(1-y)) \,,
\end{align}
In the limit of large $a\sqrt{x(1-y)}$, the asymptotic forms of the generalized hypergeometric functions can be simplified as:
\begin{align}
_2F_4(1,1;2,2,3,4; a^2 x(1-y))  & \approx  \frac{2 \sqrt{3} \, e^{3 (a \sqrt{x(1-y)})^{2/3}}}{\pi a^{16/3} (x(1-y))^{8/3}}\,,\nonumber \\
_2F_4(1,1;2,3,3,4; a^2 x(1-y))  & \approx   \frac{4 \sqrt{3} \, e^{3 (a \sqrt{x(1-y)})^{2/3}}}{\pi a^{6} (x(1-y))^{3}} \,,
\end{align}
whereas in the limit of small $a\sqrt{x(1-y)}$, both hypergeometric functions approach unity.  It is seen that the cross section in Eq.~(\ref{eq:xsq}) has an exponential growth with the exponent of $3 (a \sqrt{x(1-y)})^{2/3}$. Thus, the UHE neutrino-nucleon inelastic scattering is expected to dominate in the large $x$ and small $y$ region. From Eq.~(\ref{eq:variable definitions}),  the Bjorken variable $y$ in the laboratory frame is $1- E_{q_\nu}/E_\nu$; therefore, a small $y$ indicates $E_{q_\nu} \approx E_\nu$.  The Bjorken varaible $x$ is a parameter independent of $y$, with the theoretical range $x \in [0,1]$.  The values of $x$ and $y$ then determine the momentum transfer via $Q^2 = 2 m_N E_\nu x y$.  Thus, if $x$ is of ${\cal O}(1)$, we have $Q^2 \approx 2 m_N E_\nu y$.  With $E_\nu \sim 10^{8}$~GeV, $Q^2$ in the inelastic scattering can reach $\sim 2 \times 10^5$~GeV$^2$ if $y\sim 10^{-3}$.  Thus, when $x, 1-y\sim O(1)$, Eq.~(\ref{eq:xsq}) is only sensitive to $\exp[3 (\xi \sqrt{s}/4\pi)^{2/3}]$.  Accordingly, the cross section of $\nu N \to \nu \, \theta^n X$ is less sensitive to how the structure functions of nucleon are modeled in the scattering. 

The neutrino-nucleon inelastic scattering involves nonperturbative hadronic effects.  In the numerical analysis, we adopt the parton distribution functions (PDFs) obtained in the framework of perturbative QCD.  Due to the exponential growth feature, the resulting cross section is insensitive to the number of quark and antiquark PDFs.  For the sake of simplicity in presentation, we consider two-flavor PDFs, {\it i.e.}, $u$ and $d$ quarks and their antiquarks, while the numerical results utilize the PDFs with all active quark flavors.  We have also found that the change due to different numbers of flavors in the PDFs is immaterial.

To average the effects of proton and neutron, we take the scattered nucleon as an isoscalar, which is a combination of proton ($p$) and neutron ($n$) and denoted by $N=(n+p)/2$.  For the purpose of simplicity, we assume that the PDFs of proton and neutron have the relations:
 \begin{equation}
 u^p + u^n = d^p + d^n \equiv u + d\,,
 \end{equation}
where $q^{p(n)}$ is the quark PDF in proton (neutron), and analogous relations are applied to the antiquark PDFs.  Thus, the differential neutrino-isoscalar scattering cross section, including the hadronic effects, as a function of Bjorken variables $x$ and $y$ is obtained as:
\begin{align}
\label{eq:Xs}
\frac{d^2 \sigma_{\nu N \to \nu \, \theta^n \, X}}{dxdy} 
=& 
\frac{m^2_\nu m^4_Z G^2_F}{ 8 \pi (Q^2 + m^2_Z)^2}   \Big\{ g^2_L \left[ q(x,Q^2)  s_L(x,y) +   \bar{q}(x,Q^2) s_R(x,y)  \right] \nonumber \\ 
&  + g^2_R \left[ \bar q(x,Q^2)  s_L(x,y) +   q(x,Q^2) s_R(x,y) \right] \Big\} \,, 
 \end{align}
where $g^2_{\chi}=|g^u_\chi|^2 + |g^d_\chi|^2$ ($\chi = L,R$), and the relevant PDFs $q(x,Q^2)$ and $\bar q(x,Q^2)$  are defined as:
 \begin{align}
 q(x,Q^2) &= \frac{u(x,Q^2) + d(x,Q^2)}{2} \,, \nonumber \\
 \bar q(x,Q^2) &= \frac{\bar u(x,Q^2) +\bar d(x,Q^2)}{2}\,.
 \end{align}
It can be seen that the expression in Eq.~(\ref{eq:Xs}) is similar to that for the $\nu N\to \nu X$ process in the SM~\cite{Halzen:1984mc}, except that the functions $s_L$ and $s_R$ have a complicated dependence on $x$ and $y$ and involve an exponential factor of $\exp \left[ 3 (a \sqrt{x(1-y)}/4 \pi)^{2/3} \right]$. 
 
\subsection{Numerical analysis} 

In the following, we numerically evaluate various properties in the $\nu  N \to \nu\, \theta^n X$ inelastic scattering.  As alluded to earlier, the emission of multi-longitudinal gauge bosons is highly sensitive to the exponent of $\sqrt{s} x(1-y)$, which is associated with the off-shell neutrino invariant mass $\sqrt{q^2_\nu}=\sqrt{s z}$.  In order to show the dominant region of $\sqrt{s z}$ that contributes to the inelastic scattering, we use the variables $s_{zL,zR}$ in Eq.~(\ref{eq:szLR}) instead of $s_{L,R}$ in Eq.~(\ref{eq:Xs}).  The differential cross section as a function of $x$, $y$, and $z$ can be written as:
\begin{align}
\frac{d^3 \sigma_{\nu N \to \nu \, \theta^n \, X}}{dxdydz} 
=& \frac{m^2_\nu m^4_Z G^2_F}{ 8 \pi (Q^2 + m^2_Z)^2}   
\Big\{ g^2_L \left[  q(x,Q^2)  s_{zL}(x,y,z) +   \bar{q}(x,Q^2) s_{zR}(x,y,z)  \right] \nonumber \\ 
& + g^2_R \left[ \bar q(x,Q^2)  s_{zL}(x,y,z) +   q(x,Q^2) s_{zR}(x,y,z) \right] \Big\} \,.
\end{align}

\begin{figure}[phtb]
\begin{center}
\includegraphics[scale=0.6]{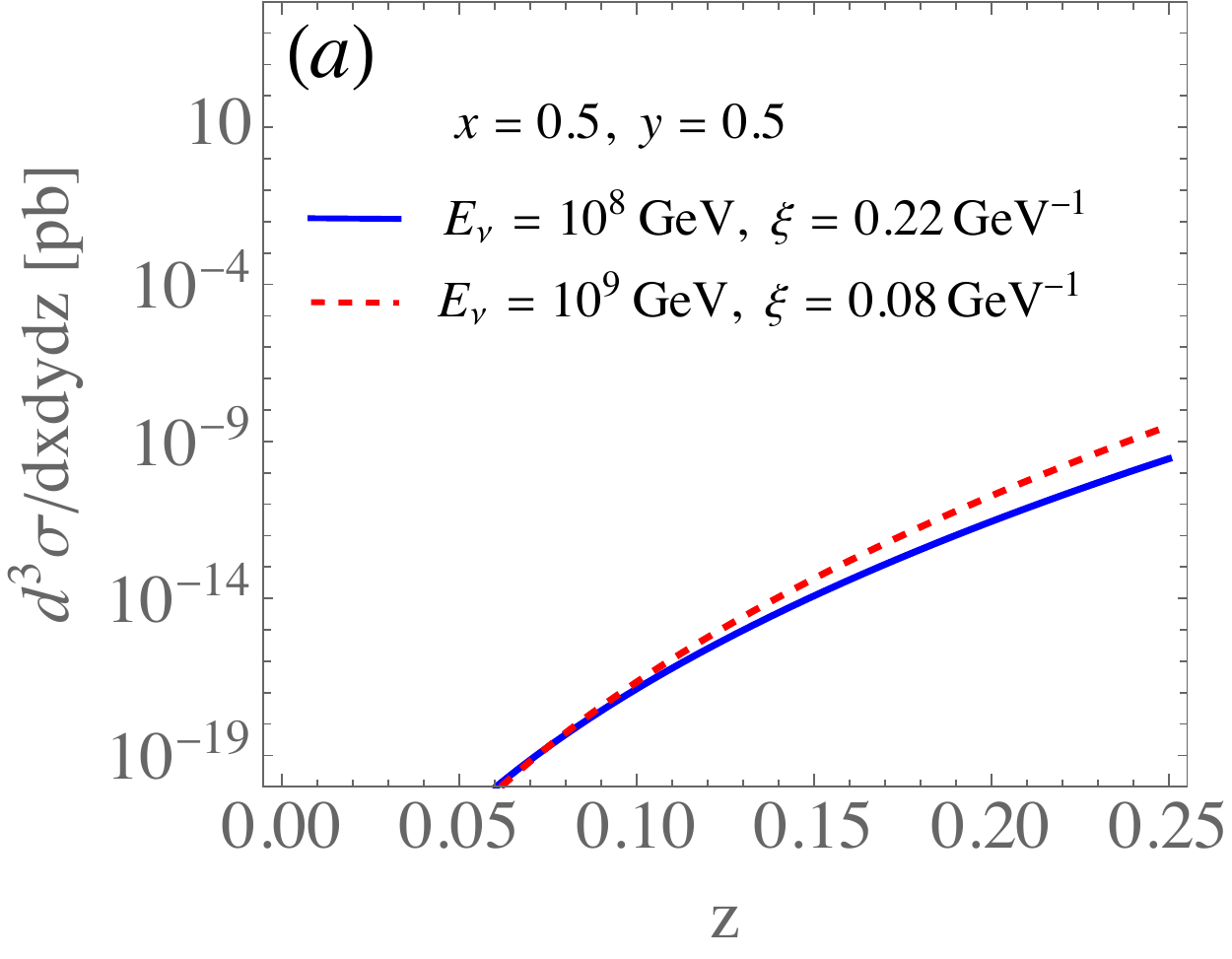}
\hspace{5mm}
\includegraphics[scale=0.6]{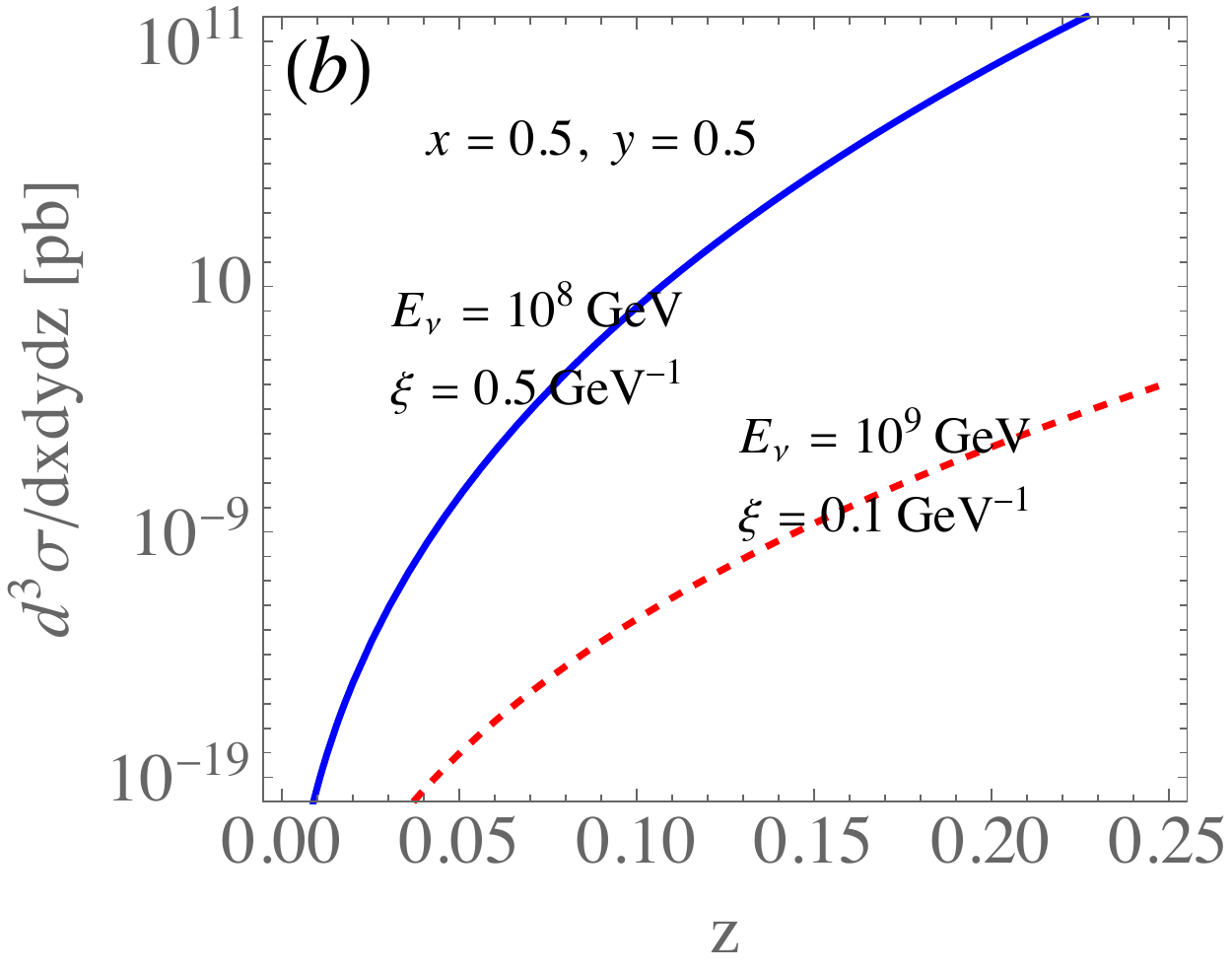}
\\
\includegraphics[scale=0.6]{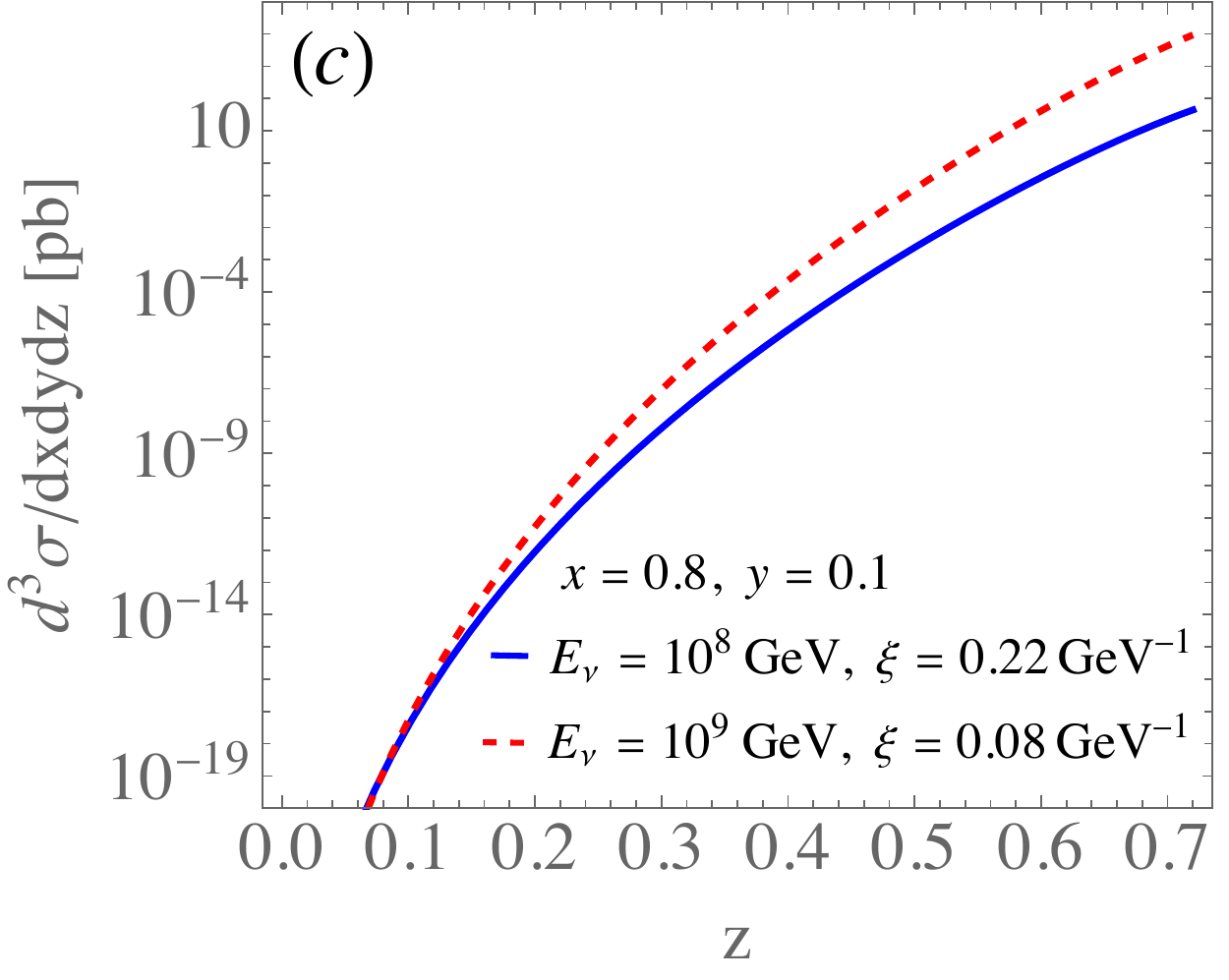}
\hspace{5mm}
\includegraphics[scale=0.6]{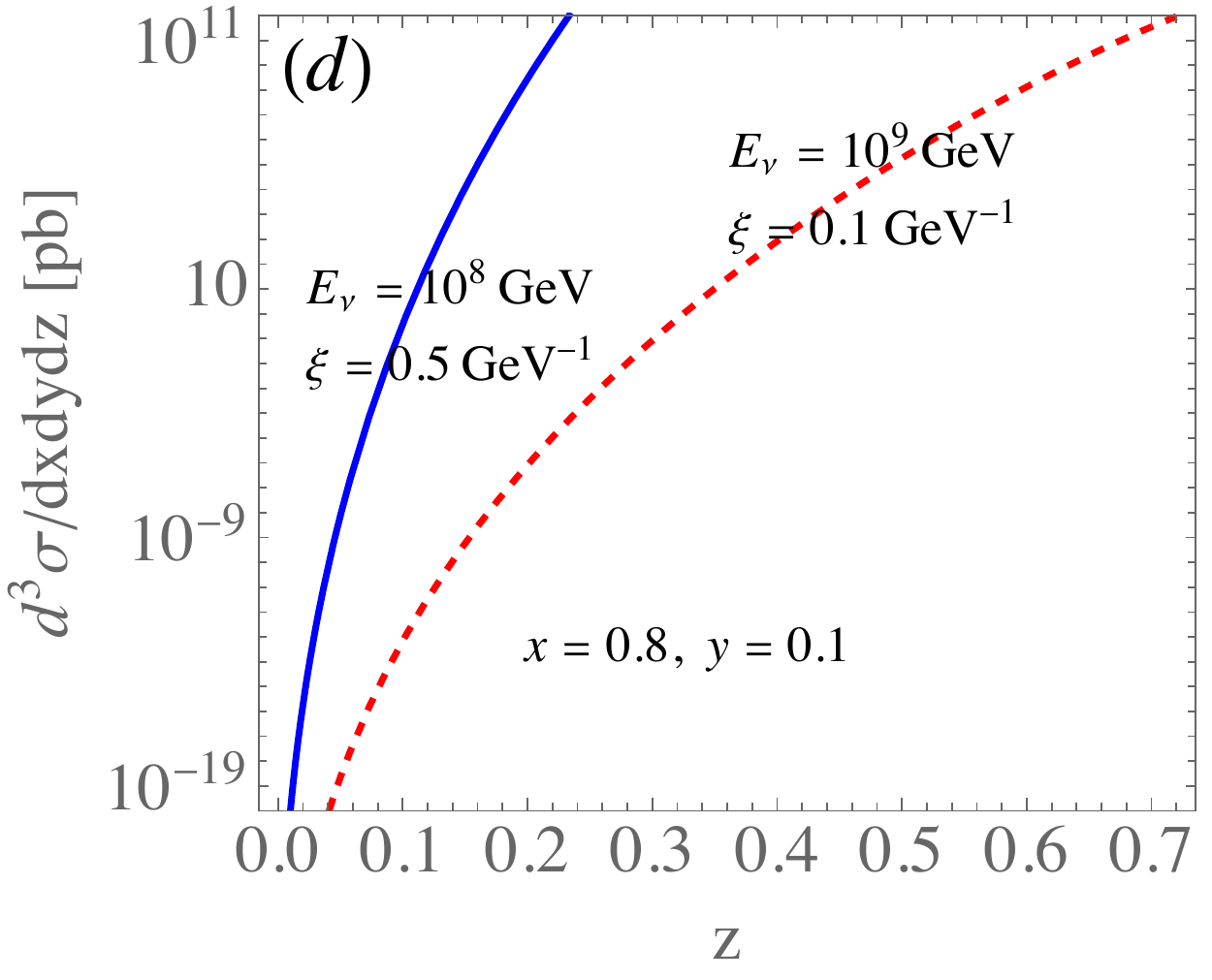}
\caption{ Partial differential cross section $d^3\sigma_{\nu N \to \nu \, \theta^n \, X}/(dx dy dz)$ as a function of $z$ when $x$ and $y$ are fixed.  In plot (a) [(b)], $x=y=0.5$ and the solid and dashed curves are for $(E_\nu/\text{GeV},\, \xi/\text{GeV}^{-1})=(10^8,\, 0.22~ [0.50])$ and $(10^9,\, 0.08~ [0.10])$, respectively.  In plot (c)[(d)], the corresponding set of $E_\nu$ and $\xi$ is used for $x=0.8$ and $y=0.1$.}
\label{fig:zs}
\end{center}
\end{figure}

Our numerical analysis is done with the help of {\tt Mathematica} and the {\tt ManeParse} package~\cite{Clark:2016jgm} using the {\tt CT10} PDFs~\cite{Lai:2010vv}.  
Since $z$ is an independent variable and its upper limit is determined by $x$ and $y$, one can thus fix the values of $x$ and $y$ to study the $z$ dependence.  We show the differential cross section as a function of $z$ in Fig.~\ref{fig:zs}, where we use $(x,y) = (0.5,0.5)$ for plots (a) and (b) and $(x,y) = (0.8,0.1)$ for plots (c) and (d).  To show the dependence on $E_\nu$ and $\xi$, we consider: $(E_\nu/\text{GeV}, \xi/\text{GeV}^{-1})=(10^8,\, 0.22)$ and $(10^9,\, 0.08)$ in plots (a) and (c), and $(E_\nu/\text{GeV}, \xi/\text{GeV}^{-1})=(10^8,\, 0.5)$ and $(10^9,\, 0.1)$ in plots (b) and (d).  It is seen that the differential cross section generally increases with $z$.  Therefore, for given $E_\nu$ and $\xi$, as shown in plots (a) [(b)] and (c) [(d)], the differential cross section is enormously enhanced when a larger $x$ and a smaller $y$ are used.  By comparing the results shown in plots (a) [(c)] and (b) [(d)], it is seen that the differential cross section is highly sensitive to the choices of $E_\nu$ and $\xi$.

It was analyzed in Ref.~\cite{Berger:2007ic} that when an UHE neutrino inelastically scattered off a nucleon in the SM, the main contribution of the Bjorken $x$ scaling variable could be estimated by:
\begin{equation}
 x_{\rm eff} \sim \frac{m^2_Z}{2 m_N E_\nu}\,.
\end{equation}
With $E_\nu\geq 10^8$~GeV, $x_{\rm eff} \lesssim 4 \times 10^{-4}$; that is, the small $x$ region dominates in the UHE neutrino-nucleon scattering.  Unlike the case in $\nu N \to \nu X$, the $\nu N \to \nu \theta^n X$ process of interest here is dominated by the large $x$ region because the exponent $\sqrt{s} x(1-y)$ approaches its maximum there, resulting in a significantly enhanced cross section.  It is also due to the exponential growth property that the cross section is highly sensitive to the $\xi$-parameter value.  Even with a slight shift in the $\xi$ value, the cross section can have an enormous change. 

\begin{figure}[phtb]
\begin{center}
\includegraphics[scale=0.6]{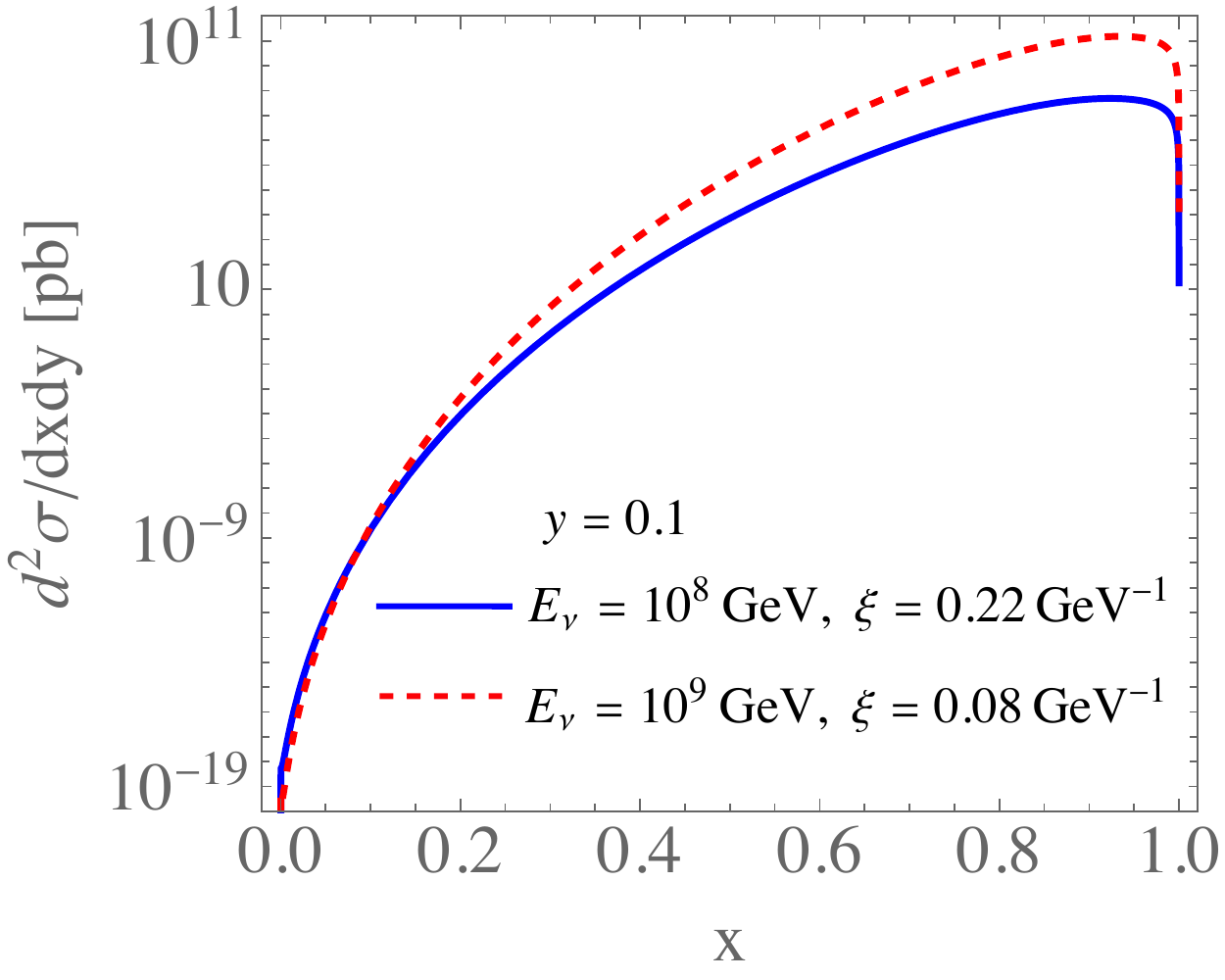}
\hspace{5mm}
\includegraphics[scale=0.6]{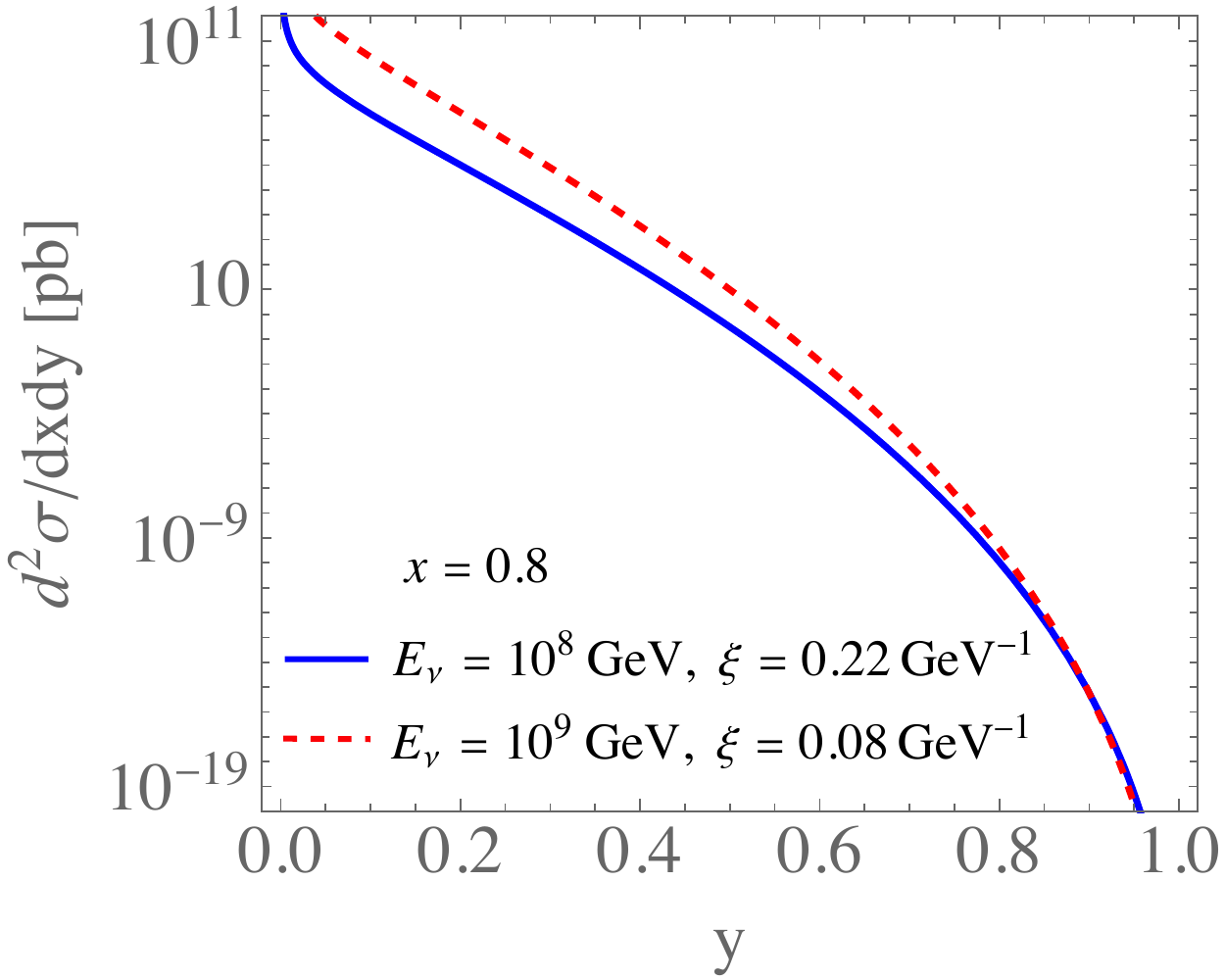}
\caption{ Differential cross section  as a function of $x$ (left plot) and $y$ (right plot), where the solid curves are for $E_\nu=10^8$~GeV and $\xi=0.22$~GeV$^{-1}$ and the dashed curves are for $E_\nu=10^9$~GeV and $\xi=0.08$~GeV$^{-1}$. We fix $y=0.1$ and $x=0.8$ in the left and right plots, respectively. }
\label{fig:diffs}
\end{center}
\end{figure}

To clearly show the individual dependence of the differential cross section on $x$ and $y$, we plot in Fig.~\ref{fig:diffs} $d^2\sigma/dx dy$ as a function of $x$ with $y=0.1$ (left plot) and of $y$ with $x=0.8$ (right plot).  The solid curves are drawn for $E_\nu=10^8$~GeV and $\xi=0.22$~GeV$^{-1}$, and the dashed curves are for $E_\nu=10^9$~GeV and $\xi=0.08$~GeV$^{-1}$.  The choices of these two sets of parameters are made so that the total cross section of $\nu N \to \nu\, \theta^n X$ is slightly less than that of $\nu N \to \nu\, X$.  It is seen that the differential cross section increases with $x$ for most of the region, whereas its dependence on $y$ is monotonically decreasing.  The sudden drop of the differential cross section for $x\approx 1$ in the left plot is the result of vanishing PDFs when $x\to1$.  The rapidly decreasing behavior when $y$ gets close to unity in the right plot mainly comes from the vanishing exponent of $\sqrt{s} x(1-y)$.

\begin{figure}[phtb]
\begin{center}
\includegraphics[scale=0.45]{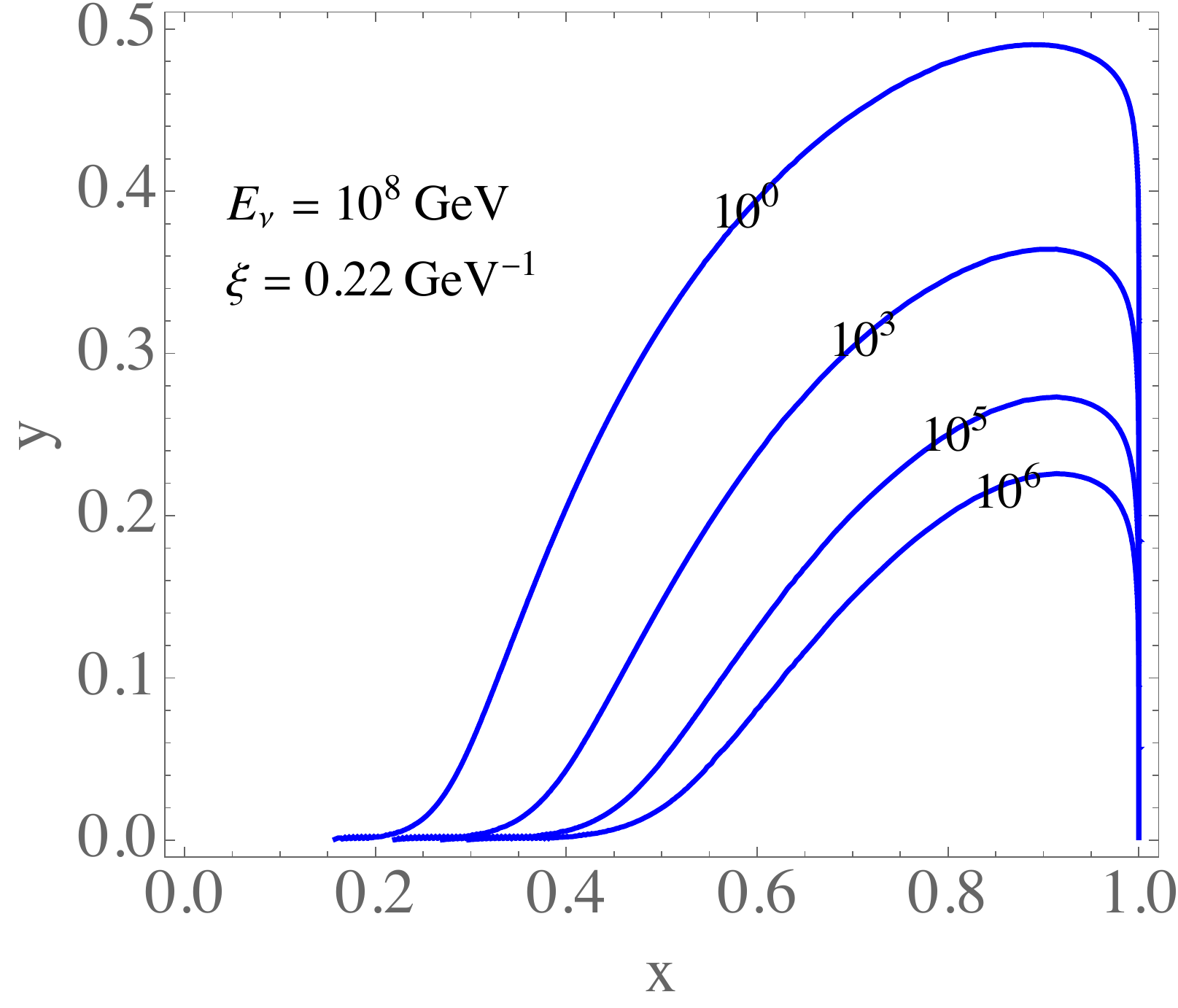}
\hspace{5mm}
\includegraphics[scale=0.45]{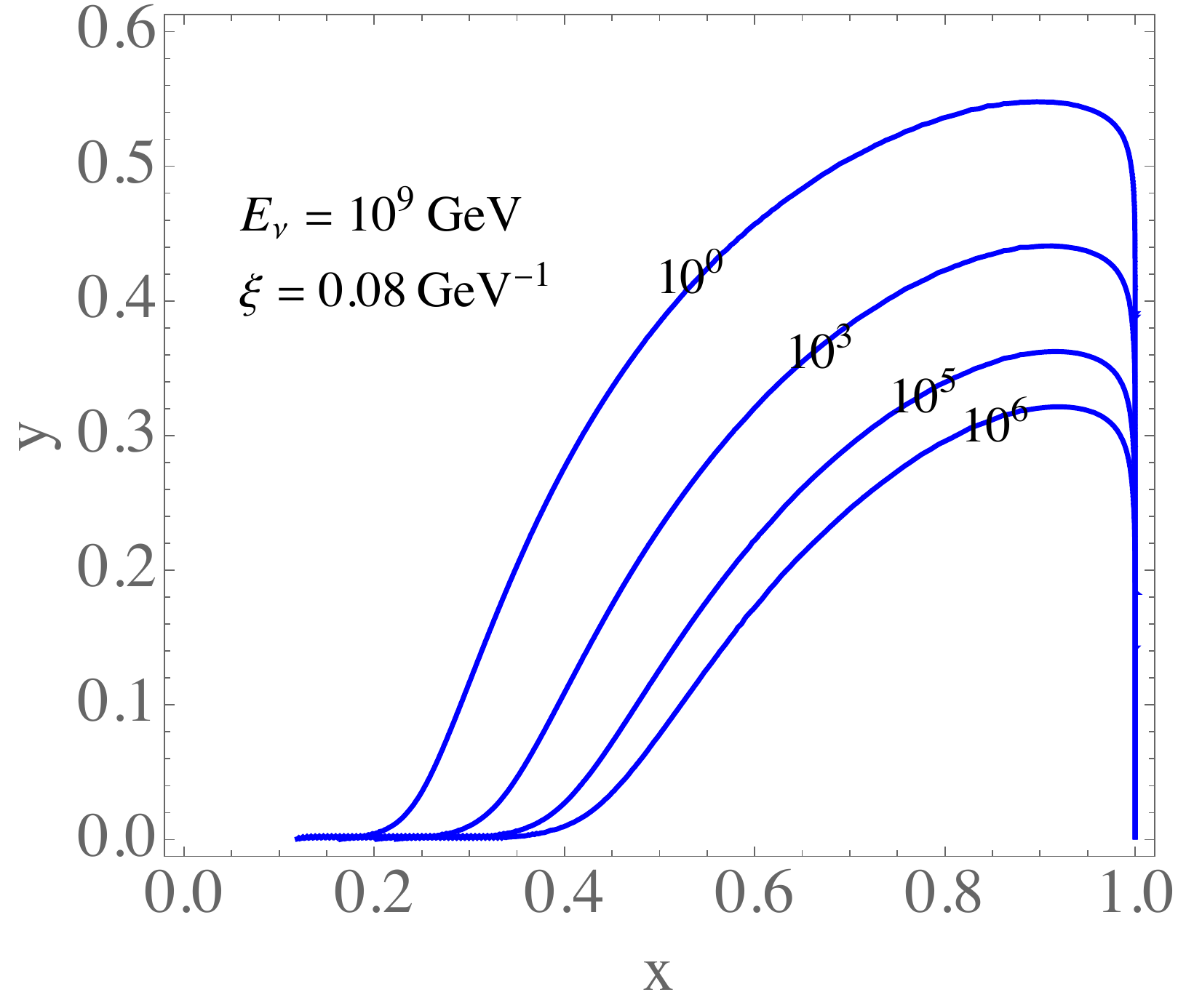}
\caption{Contours of the differential cross section, in units of pb, with respect to the Bjorken variables $x$ and $y$, where $E_\nu=10^8$ GeV and $\xi=0.22$ GeV$^{-1}$ are used in the left plot and $E_\nu=10^9$ GeV and $\xi=0.08$ GeV$^{-1}$ are used in the right plot.}
\label{fig:Contours}
\end{center}
\end{figure}

In addition, we show the contours of $d^2\sigma/dxdy$ in the plane of $x$ and $y$ in Fig.~\ref{fig:Contours}, where the left and right plots correspond to the cases of $E_\nu=10^8$~GeV with $\xi=0.22$~GeV$^{-1}$ and $E_\nu=10^9$~GeV with $\xi=0.08$~GeV$^{-1}$, respectively.  If it is to be of the same order of magnitude as but slightly smaller than the SM cross section of $\nu N \to \nu X$, $d^2\sigma/dxdy$ should fall in the range of $O(10^2-10^3)$~pb, corresponding to a small region in the $x$-$y$ plane and reflecting the effect of the $\exp\left[ 3 (a \sqrt{x(1-y)}/4 \pi)^{2/3} \right]$ factor.
  
Next, we calculate the total cross section of the UHE neutrino-nucleon inelastic scattering by integrating out the $x$ and $y$ variables.  Since the $\nu N \to \nu\, \theta^n X$ process with multiple close-to-massless Goldstone boson emissions cannot be distinguished from $\nu N \to \nu X$ in practical measurements, their contributions should be added together to compare with experimental data.  In Table~\ref{tab:Xs}, we list the SM predictions on $\sigma(\nu N \to \nu X)$ for several values of $E_\nu$.  For fixed values of $E_\nu$, the only free parameter in Eq.~(\ref{eq:Xs}) is $\xi$.  According to the IceCube measurements~\cite{IceCube:2020rnc}, the measured $\nu$-$N$ scattering cross sections via both charged and neutral currents are consistent with the SM results up to $E_\nu \sim 10^6$~GeV.  We will thus use their finding to constrain the parameter $\xi$ in the anomalous $U(1)$ EFT. 

\begin{figure}[phtb]
\begin{center}
\includegraphics[scale=0.45]{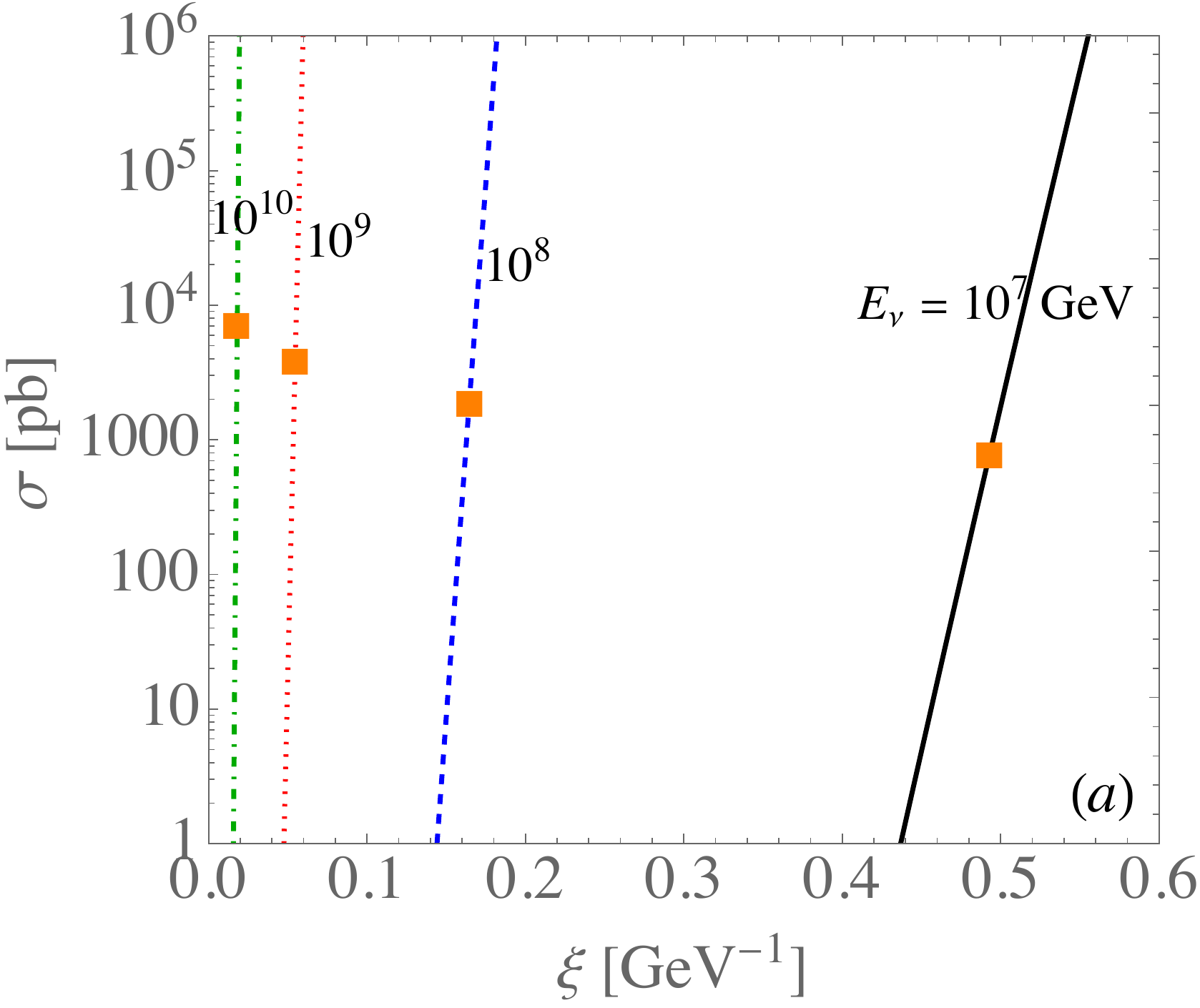}
\hspace{5mm}
\includegraphics[scale=0.45]{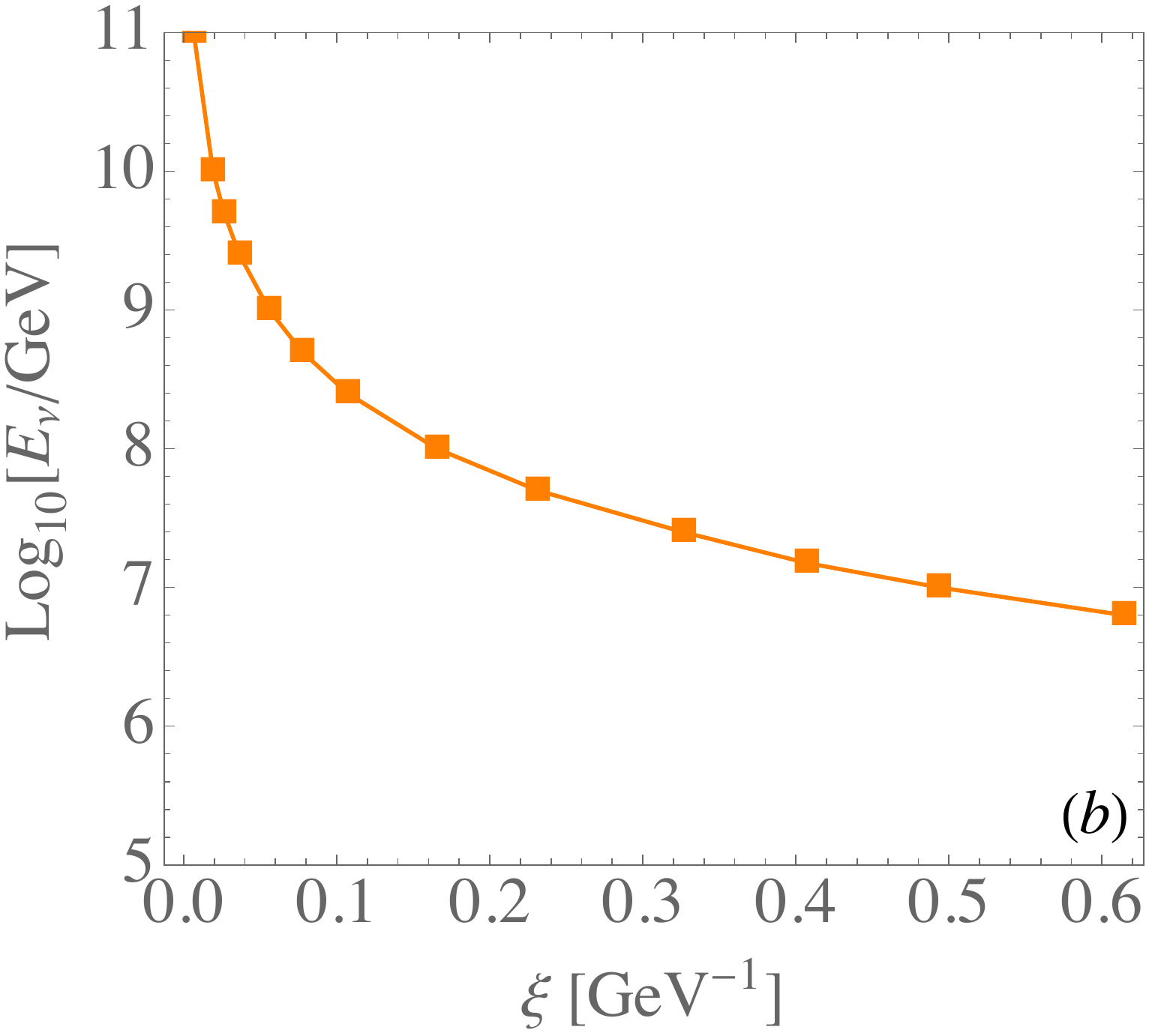}
\caption{ (a) Cross section of the $\nu N \to \nu\, \theta^n X$ process with different incident neutrino energies. (b) Upper limit of $\xi$, where  the squares denote the SM predictions for the cross section of $\nu N\to \nu X$ at the corresponding $E_\nu$~\cite{Block:2010ud}.}\label{fig:Xs}
\end{center}
\end{figure}

To numerically illustrate the correlation between the inelastic scattering cross section $\sigma(\nu N \to \nu \, \theta^n X)$ and $\xi$, we take some benchmark values of $\xi$ such that the resulting $\sigma(\nu N \to \nu \, \theta^n X)$ is slightly smaller than the SM prediction, as given in the third and fourth columns of Table~\ref{tab:Xs}.  The numerical results have twofold implications: (i) If any excess is observed at IceCube in the future, the new interaction of neutrino with multi-longitudinal gauge bosons based on the $U(1)$ anomalous EFT can be a candidate mechanism to explain the anomaly; and (ii) the IceCube measurement can give a more stringent bound on $\xi$ if they can probe neutrinos of higher energies.  For example, the constraint of $\xi=0.612$ from $E_\nu=6.3 \times 10^{6}$~GeV is stronger than $\xi=0.769$ from $pp\to W^* \to \ell \bar\nu$ as given in Ref.~\cite{Ekhterachian:2021rkx}.   Moreover, we show $\sigma(\nu N\to \nu\, \theta^n X)$ as a function of $\xi$ in Fig.~\ref{fig:Xs}(a) for different neutrino energies $E_\nu=6.3 \times 10^{6}-10^{10}$~GeV.  As a reference, we also show in orange squares the SM values of $\sigma(\nu N \to \nu X)$ at the corresponding $E_\nu$.  Due to the lack of data at higher energies at the moment, we shall take the SM results as the reference upper bounds for $\nu N \to \nu \theta^n X$.  The upper limit of $\xi$ for the corresponding $E_\nu$ can be read off from Fig.~\ref{fig:Xs}(b), where the SM results are estimated by following the formulas given in Refs.~\cite{Berger:2007ic,Block:2010ud}.  The fact that the lines become steeper for higher neutrino energies is again attributed to the exponential growth feature in the UHE neutrino-nucleon inelastic scattering.

\begin{table}[htbp]
  \centering
  \begin{tabular}{cccc} \\  \hline \hline  
   $E_\nu$ (GeV) ~~& ~~ $\sigma_{\rm SM}$ (cm$^2$) ~~& ~~ $\sigma_{\nu N \to \nu \theta^n X}$ (cm$^2$) ~~& ~~$\xi$ (GeV$^{-1}$) \\  \hline 
   $6.3 \times 10^{6}$ & $6.27 \times 10^{-34}$ & $5.29 \times 10^{-34}$  & 0.612 \\  
   $10^{7}$ & $7.55 \times 10^{-34}$ & $6.11\times 10^{-34}$ & 0.491 \\ 
   $10^{8}$ & $1.82 \times 10^{-33}$ &  $1.07\times 10^{-33} $ & 0.162   \\  
    $10^9$ & $3.76 \times 10^{-33}$  & $1.78\times 10^{-33}$ & 0.054 \\
    $10^{10}$ & $6.89 \times 10^{-33}$  &  $4.86\times 10^{-33} $ &  0.018 \\ \hline \hline  
  \end{tabular}
  \caption{Cross sections of the $\nu N \to \nu X$ process in the SM, quoted from Ref.~\cite{Block:2010ud}, and the $\nu N \to \nu \theta^n X$ process for different values of the neutrino energy $E_\nu$.  The latter is evaluated using different benchmark values of $\xi$ such that it does not exceed the corresponding cross section of the former.}
\label{tab:Xs}
\end{table}

\section{Summary \label{sec:Summary}}

We have studied an anomalous $U(1)$ effective field theory with an extremely light massive gauge boson at low energies, where the gauge anomaly arises from the right-handed neutrino and the gauge invariance is broken by the Dirac neutrino mass.  When the longitudinal component of the gauge boson is nonlinearly represented by a Goldstone or Stueckelberg field and couples to the neutrino, it is possible for neutrinos to emit multi-longitudinal gauge bosons and the associated amplitude generally features in an exponential growth with energy.

We thus considered the ultra-high-energy neutrino-nucleon inelastic scattering, the process occurring in the IceCube experiment.  It has been found that the inelastic cross section of $\nu N \to \nu \theta^n X$ is sensitive to the parameter $g_X/m_X$.  When taking the recent IceCube data for $\nu N \to \nu X$ as the upper bound on the cross section of $\nu N \to \nu \theta^n X$, we obtained $g_X/m_X < 0.612$, stricter than that obtained from $pp\to W^* \to \ell \bar\nu$ at $2$~TeV.  We thus conclude that ultra-high-energy neutrino scattering off nucleons can serve as a better process to uncover/constrain the model.

{
It is worth mentioning that in addition to the IceCube experiment, the proposed next-generation UHE neutrino experiments, such as Giant Radio Array for Neutrino Detection (GRAND)~\cite{GRAND:2018iaj} and Probe Of Extreme Multi-Messenger Astrophysics (POEMMA)~\cite{AlvesBatista:2018zui}, can probe the $\tau$ neutrino energy up to the order of $10^9$~GeV.  The expected precision in the $\nu_\tau$-$N$ scattering cross section as extracted from these experiments had been studied in Ref.~\cite{Denton:2020jft}. }

Finally, we make some remarks on other types of colliders than hadron colliders that can potentially probe the novel exponential growth effect.  Even though an International linear collider (ILC) with $500$~GeV or $1$~TeV after energy upgrade can produce high-energy off-shell $Z$-bosons~\cite{Baer:2013cma,vanderKolk:2016akp}, the constraint on $g_X/m_X$ through $e^+ e^- \to Z^* \to \nu \bar\nu \theta^n$ at even $\sqrt{s}=1$~TeV is weaker than that from $pp\to W^*\to \ell \bar\nu $ at the LHC because of the similar coupling structure.  The proposed Compact Linear Collider with $\sqrt{s}=3$~TeV could be better than the ILC; however, a similar bound from the LHC with higher integrated luminosity at $\sqrt{s}=14$~TeV may already be available~\cite{CMS:2021pmn}.  The most promising machine is the muon collider, where the initial energy can reach $10$~TeV~\cite{Delahaye:2019omf,Long:2020wfp}.  Applying the estimate used in Ref.~\cite{Ekhterachian:2021rkx}, the bound at such an energy scale is evaluated approximately as $g_X/m_X \lesssim  0.2$.  

\section*{Acknowledgments}

CWC would like to thank Yuhsin Tsai for useful discussions of their earlier work.  This work was supported in part by the Ministry of Science and Technology, Taiwan under the Grant Nos.~MOST-110-2112-M-006-010-MY2 and MOST-108-2112-M-002-005-MY3.


\end{document}